\documentclass[printer]{gPHT2e}
\newcommand{\BEQ}{\begin{equation}}     
\newcommand{\BEA}{\begin{eqnarray}}
\newcommand{\EEQ}{\end{equation}}       
\newcommand{\EEA}{\end{eqnarray}}
\def\be{\begin{equation}}
\def\ee{\end{equation}}
\def\bc{\begin{center}}
\def\ec{\end{center}}
\newcommand{\D}{{\rm d}}

\input epsf.sty

\begin{document}

\doi{10.1080/014115940xxxxxxxxxxxx}
 \issn{1029-0338}
\issnp{0141-1594} \jvol{00} \jnum{00} \jyear{2005}
\jmonth{January}

\title{Microcanonical analysis of small systems}
\author{Michel Pleimling$^{1}$ and Hans Behringer$^{1,2}$}

\received{\today}

\maketitle

\affil{$^1$ Institut f\"ur Theoretische Physik I,
Universit\"at Erlangen-N\"urnberg, D -- 91058 Erlangen, Germany}

\affil{$^2$ Fakult\"at f\"ur Physik, Universit\"at Bielefeld, D -- 33615
Bielefeld, Germany}

\begin{abstract}
The basic quantity for the description of the statistical
properties of physical systems is the density of states or
equivalently the microcanonical entropy. Macroscopic quantities of
a system in equilibrium can be computed directly from the
entropy. Response functions such as the susceptibility
are for example related to the curvature of the
entropy surface. Interestingly,
physical quantities in the microcanonical ensemble show
characteristic properties of phase transitions already in 
finite systems. In this paper we investigate these characteristics 
for finite Ising systems. The
singularities in microcanonical quantities which announce a
continuous phase transition in the infinite system are characterised by
classical critical exponents. Estimates of the non-classical
exponents which emerge only in the thermodynamic limit can nevertheless be
obtained by analyzing effective exponents or by applying a microcanonical
finite-size scaling theory. This is explicitly demonstrated for two-
and three-dimensional Ising systems.

\end{abstract}

\section{Introduction}
In the study of the static properties of finite systems the main quantity
of interest is the density of states $\Omega$ as it contains the complete 
physical information of the investigated system \cite{Gro01b}. To be specific, consider
a magnetic system where $\Omega = \Omega(E,M)$ is a function of the energy $E$
and of the magnetization $M$ (which can be a scalar or a vector quantity, depending
on the system). With the density of states at hand one may proceed
in the canonical way by computing the free energy from which other 
quantities such as the susceptibility of the specific heat can be derived.
In the last years, however, an increasing number of groups have chosen
to directly investigate the density of states or, equivalently, the
microcanonical entropy $S =  \ln \Omega$ (setting $k_B=1$).
The origin of this increasing interest in the microcanonical approach
can be found in the observation that ensembles are not always equivalent
\cite{Gro01b,thirr70,bixon89,Hue94a,elli00,daux00,Bar01,Isp01,gulm02}.
The inequivalenve of ensemble is for example encountered in situations
where the macroscopic limit (for example due to the presence of long-range
interactions such as the gravitational force) does not exist.
But also in finite systems with short-range interactions the canonical
and the microcanonical ensembles may give different results although
the ensembles have to become equivalent in the infinite volume limit. This is
especially the case for finite systems which in the macroscopic limit
undergo a phase transition. It is the purpose of this paper to discuss 
the finite-size signatures of phase transitions in the microcanonical
ensemble \cite{kast00,Gro00b,huel02,plei04,hove04,behr05,rich05}.

The paper is organized in the following way. In Section 2 we recall
that in the microcanonical approach all quantities of interest
can exclusively be expressed as derivatives of the entropy \cite{Gro01b}. In
Section 3 we briefly discuss the case of discontinuous phase
transitions which lead to a characteristic behaviour of the
specific heat in finite microcanonical systems. Section 4 is the 
main part of this paper. It is devoted to the signatures of
continuous phase transitions encountered in the microcanonical
analysis of small systems. We also discuss different ways for
extracting the values of critical exponents 
directly from the derivatives of the microcanonical entropy. Conclusions
are given in Section 5.

\section{Microcanonically defined quantities}

We use in the following the language of magnetism and consider a magnetic system
with $N$ degrees of freedom isolated from any environment. The state of the
system is then characterized by its energy $E$ and its magnetization $M$.

In the microcanonical analysis one considers the microcanonical entropy 
density
\begin{equation} \label{eq:s}
s(e,m) = \frac{1}{N} S(eN,mM) = \frac{1}{N} \ln \Omega (eN,mM) 
\end{equation}
as function of the energy density $e = E/N$ and the magnetization density
$m=M/N$. The entropy surface $s(e,m)$ still exhibits a weak size
dependence which is not explicitly denoted here. Sometimes one is only interested in the energy dependence
and therefore studies the reduced specific entropy
\begin{equation} \label{eq:sR}
s_R(e)= \frac{1}{N} \ln \Omega_R (eN)
\end{equation}
with 
\begin{equation}\label{eq:omegaR}
\Omega_R (E) = \sum\limits_M \Omega (E,M).
\end{equation}
Microcanonical quantities can be expressed by  functions of derivatives
of the specific microcanonical entropy. For example, the microcanonical
temperature reads
\begin{equation} \label{eq:T}
T = \left( \frac{\D s_R(e)}{\D e} \right)^{-1},
\end{equation}
which leads to the following expression for the specific heat:
\begin{equation} \label{eq:c}
c(e)= - \frac{\left( \frac{\D s_R(e)}{\D e} \right)^2}{\frac{\D^2s_R(e)}{\D e^2}}.
\end{equation}
It is worth noting that the microcanonical temperature and specific heat
can also be defined through the entropy density (\ref{eq:s}). These differently
defined quantities get identical in the infinite volume limit. We refer the
reader to \cite{behr05} for a thorough discussion of this point.

Further quantities discussed in the following include the spontaneous
magnetization and the susceptibility. The spontaneous magnetization $m_{sp,N}(e)$ is
defined to be the value of $m$ where the entropy (\ref{eq:s}) at a fixed value of $e$
has its maximum with respect to $m$ \cite{kast00,huel02,behr05,behr04}:
\begin{equation} \label{eq:msp}
m_{sp,N}(e) :\Longleftrightarrow s(e, m_{sp,N}(e)) = \max_m s(e,m).
\end{equation}
To better understand this definition consider the density
of states $\Omega(eN,mN)= \exp (N s(e,m))$ for a given energy $e$. The density of
states exhibits a sharp maximum as the overwhelming majority of accessible states belongs
to the value $m_{sp,N}$ of $m$ where $s$ has its maximum. This is the reason 
why we identify $m_{sp,N}$ with
the spontaneous magnetization of the finite system.
This definition of the microcanonical spontaneous magnetization ensures that we recover the
canonically defined spontaneous magnetization in the macroscopic limit.
Note also that one can define a magnetic field for the
microcanonical ensemble which appears as the conjugate variable to
$m$. At the spontaneous magnetization $m_{sp,N}$ the
associated magnetic field becomes zero, see \cite{kast00} for further details.

Of further interest is the suceptibility \cite{kast00}
\begin{equation} \label{eq:chi}
\chi(e,m)=- \frac{\frac{\partial s}{\partial e} \frac{\partial^2 s}{\partial e^2}}{\frac{
\partial^2 s}{\partial e^2} \frac{\partial^2 s}{\partial m^2} - \left(
\frac{\partial^2 s}{\partial e \partial m} \right)^2 }
\end{equation}
where the denominator on the right hand side is in fact the local curvature of
the entropy surface. As it stands, Eq. (\ref{eq:chi}) is only valid in cases where
the order parameter is a scalar. For an order parameter with vector character
(as encountered for example in classical spin models like the $XY$ or the
Heisenberg model) the susceptibility is given by a matrix
\cite{behr04b,rich05}. The entries of the susceptibility matrix are
then dependent on the local curvature with respect to the different order
parameter components.

Closing this Section, we reemphasize that the microcanically defined
quantities all fulfill the important request that they become identical to
the better known canonical quantities in the infinite system after the
trivial transformation from the natural variables $e$ and $m$ of the
microcanonical description to the natural variables temperature $T$
and magnetic field $h$ of the canonical approach.

\section{Discontinuous phase transitions}
The interest in the microcanonical analysis really started with the observation
that discontinuous phase transitions display typical and very
interesting finite-size signatures in the microcanonical treatment \cite{bixon89,Hue94a,Gro01b}. 
Indeed, a discontinuous phase transition
reveals itself in finite systems by a convex intruder in the microcanonical
entropy which originates from states of coexistence of different phases.
In the canonical treatment these states are unstable and cannot be accessed by
equilibrium methods. Interestingly, this convex intruder leads to
a characteristic backbending of the caloric curve and thus to a negative specific
heat. The appearance of a negative specific heat may look like an oddity of
the microcanonical approach. However, there have been recent claims that 
negative specific heats had been measured (even so in a somewhat indirect way)
in experiments on nuclear fragmentation \cite{Dag00} or on the solid-liquid transition
of atomic clusters \cite{Sch01a}.

\section{Continuous phase transitions}
The main intention of this paper is to discuss the typical finite-size
signatures of continuous phase transitions encountered in the microcanonical
approach. We thereby focus on classical spin systems and shall discuss in the
following the two- and three-dimensional Ising models on hypercubic
lattices with periodic boundary conditions. The Ising model is defined by the Hamiltonian
\begin{equation} \label{eq:ising}
{\mathcal H} = - \sum\limits_{\langle i,j \rangle} \, S_i S_j
\end{equation}
where the lattice site $i$ is characterized by the classical variable $S_i = \pm 1$.
The sum in Equation (\ref{eq:ising})
is over nearest neighbour pairs. The two-dimensional three-state
Potts \cite{behr04,hove04,behr05} and the three-dimensional $XY$ \cite{rich05} models have also been studied 
recently in the present context, but we will
not discuss these models here. All the mentioned models have in common that the
corresponding infinite system displays a continuous phase transition separating
a paramagnetic phase at high temperatures (energies) from a ferromagnetically
ordered phase at low temperatures (energies).
\subsection{Signatures in small systems}
In the canonical ensemble the typical features of spontaneous symmetry breaking
like the abrupt onset of the order parameter or a diverging susceptibility
are only encountered in the infinite system. In any finite system one does not
observe diverging quantities but only rounded maxima. This is, however,
completely different in the microcanonical ensemble where the features
of spontaneous symmetry breaking turn up already for finite systems
\cite{kast00,huel02,behr04,plei04,behr05,rich05,behr03}.

\begin{figure}
\centerline{\epsfxsize=6.0in\ \epsfbox{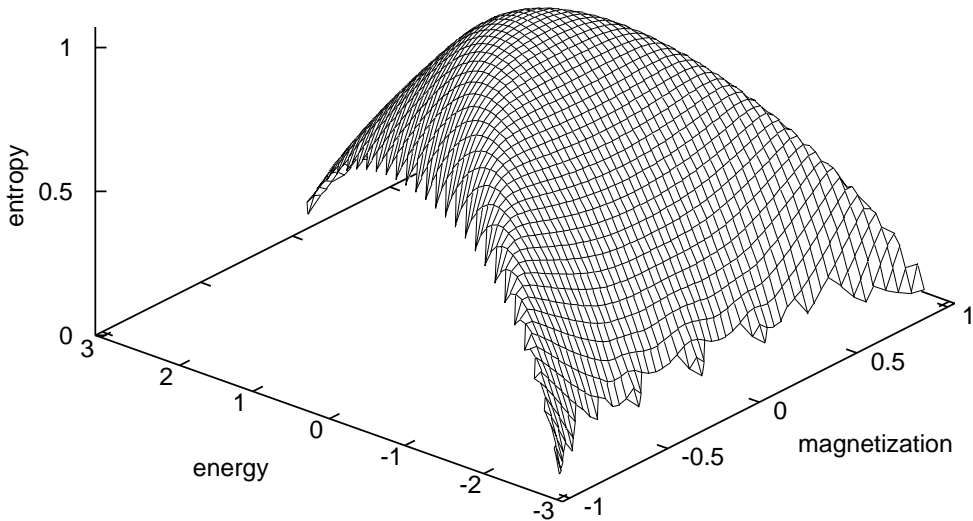}} 
\caption{Entropy surface $s(e,m)$ of the three-dimensional Ising model with
$N=6^3$ spins as function of the energy $e$ and of the
magnetization $m$.} \label{fig_1}
\end{figure}

\begin{figure}
\centerline{\epsfxsize=4.0in\ \epsfbox{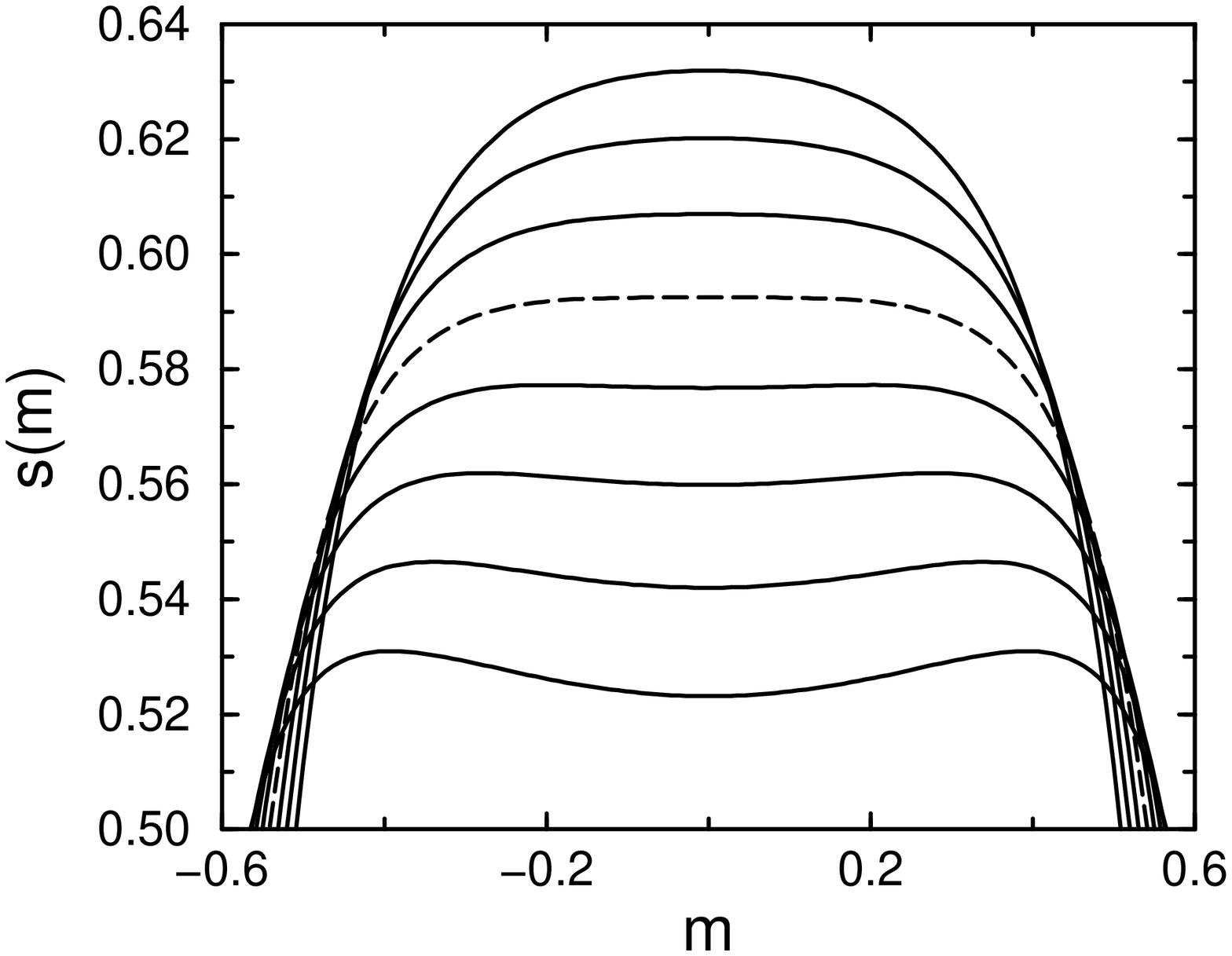}} 
\caption{Cut through the entropy surface at various fixed values of $e$. For $e \geq e_{c,N}$
the entropy exhibits a maximum at $m=0$. For $e < e_{c,N}$, two maxima
at nonzero magnetizations are visible. At $e = e_{c,N}$ (dashed line) the extremum
at $m=0$ changes from a maximum at higher energies to a minimum at
lower energies so that the curvature changes its sign.} \label{fig_2}
\end{figure}

It is instructive to first inspect directly the entropy surface $s(e,m)$.
Figure \ref{fig_1} displays this entropy surface for a three-dimensional 
Ising model with $6^3$ spins. The fully magnetized states at $e=-3$ and $m= \pm 1$
form the two degenerate gound states, whereas the fully disordered macrostate
at $e=m=0$ has the highest degeneracy of all. This entropy surface has been
computed by a very efficient numerical method based on the concept of
transition variables \cite{huel02}. More insights are gained by looking at cuts through
the entropy surface along the magnetization at a fixed energy $e$, see Figure \ref{fig_2}. 
One observes that the entropy exhibits
only one maximum at $m=0$ for energies $e$ above an energy $ e_{c,N}$, whereas for energies $e < e_{c,N}$ two
maxima are seen at nonzero magnetizations, separated by
a minimum at $m=0$. At the energy $e_{c,N}$ the curvature in
magnetization direction vanishes at $m=0$.

\begin{figure}
\centerline{\epsfxsize=4.0in\ \epsfbox{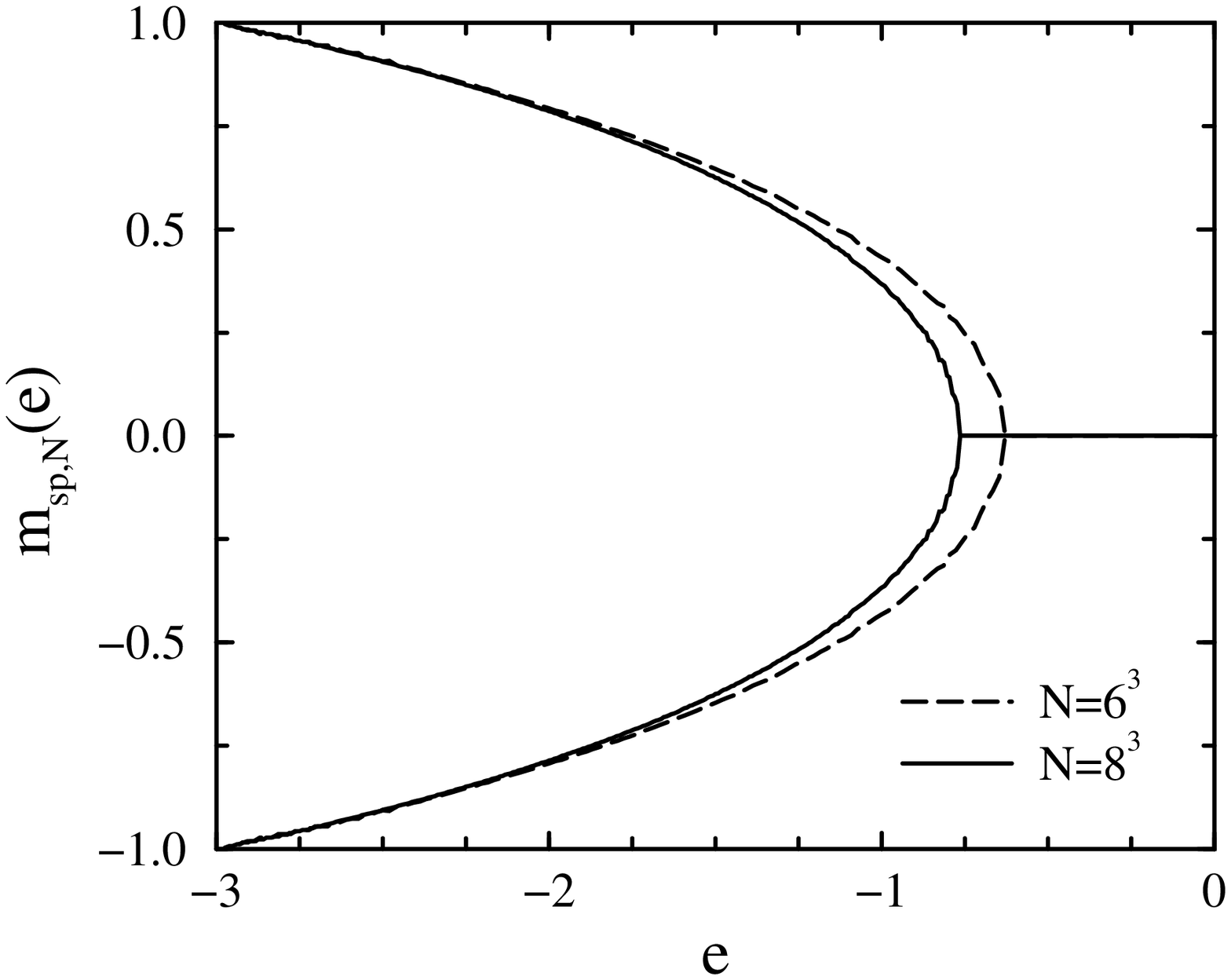}} 
\caption{Spontaneous magnetization $m_{sp,N}$ as function of the energy $e$ 
measured in three-dimensional Ising models with $6^3$ and $8^3$ spins.
At the size-dependent energy $e_{c,N}$ a sharp onset of the order parameter is
observed.} \label{fig_3}
\end{figure}

Recalling that the spontaneous magnetization is identified with
the magnetization where for a fixed energy the density of states is the largest,
one can directly extract the value of the spontaneous magnetization at a 
given energy from cuts like those shown in Figure \ref{fig_2}.
The resulting behaviour of the spontaneous magnetization
is discussed in Figure \ref{fig_3} 
for three-dimensional Ising models containing $6^3$ and
$8^3$ spins. Two different regimes are observed: for $e \geq e_{c,N}$
the spontaneous magnetization is zero, whereas for $e < e_{c,N}$
it is given by $m = \pm m_{sp,N}$, in accordance with the appearance
of two maxima of the entropy surface at lower energies.
At the well-defined size-dependent energy $e_{c,N}$ the order
parameter sets in abruptly already in finite systems. The energy
$e_{c,N}$ may therefore be called transition energy.

\begin{figure}
\centerline{\epsfxsize=4.0in\ \epsfbox{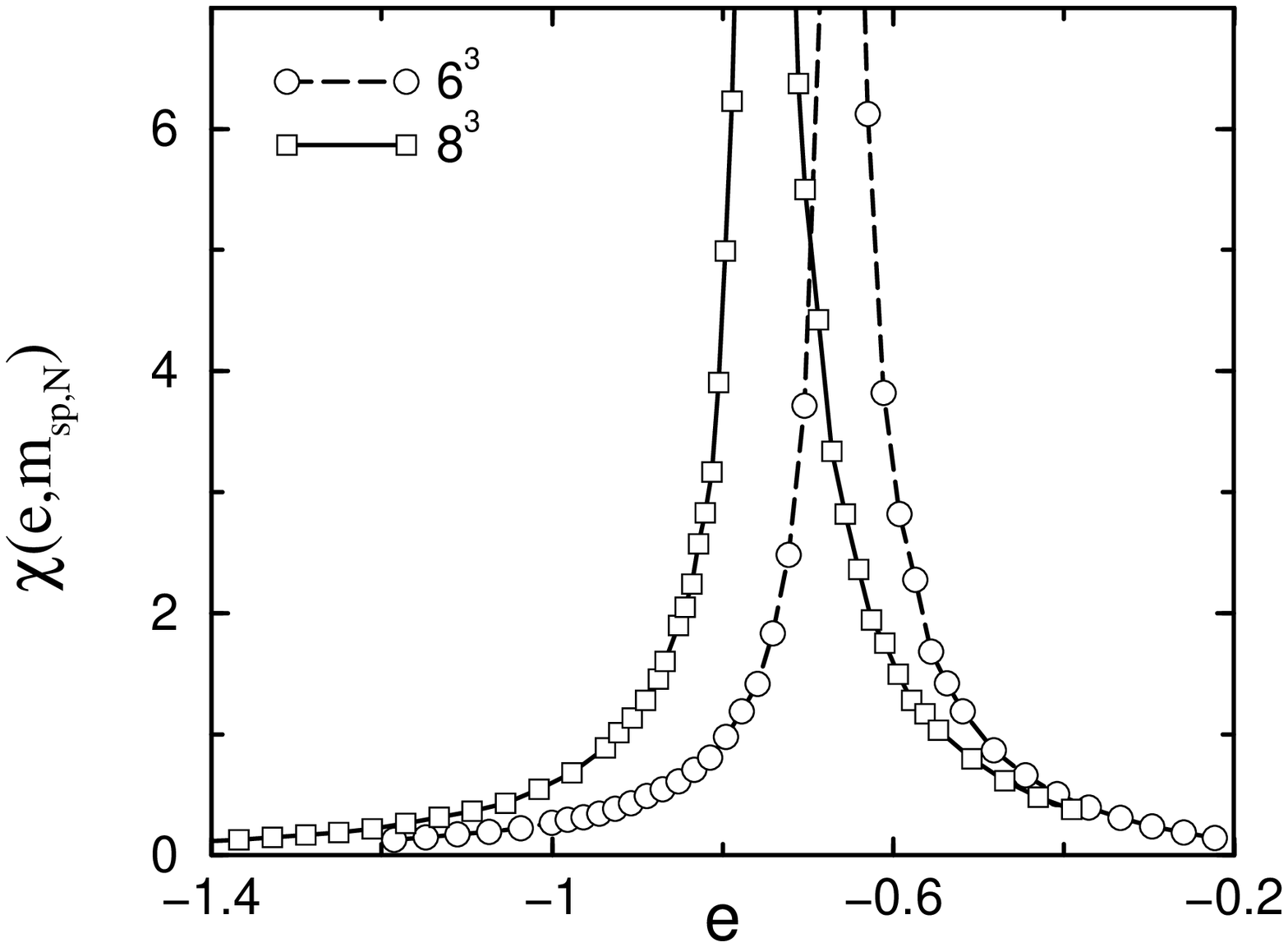}} 
\caption{Magnetic susceptibility $\chi(e,m_{sp,N})$ as function of the energy $e$ 
measured in three-dimensional Ising models with $6^3$ and $8^3$ spins.
The microcanonically defined reponse function diverges at $e_{c,N}$ already
in finite systems.} \label{fig_4}
\end{figure}

Another remarkable finite-size signature of continuous phase transitions is
shown in Figure \ref{fig_4}. A divergence of the magnetic susceptibility is
observed for finite systems at the same energy $e_{c,N}$ at which the
order parameter bifurcates. This divergence is in fact readily understood
by recalling, see Equation (\ref{eq:chi}), that the susceptibility is proportional
to the inverse of the curvature of the entropy surface. It is the existence
of a point of vanishing curvature that is responsible for the observed
divergence of the susceptibility, see Figure \ref{fig_2}.

The microcanonical specific heat, on the other hand, does not diverge in finite 
systems. 
It nevertheless also displays an intriguing finite-size behaviour,
as revealed recently for models with a diverging specific heat in 
the  infinite system \cite{behr05}. Indeed, in the microcanonical ensemble
the maximum of the specific heat of the corresponding finite systems varies non-monotonically  
when increasing the size of the system. Starting with small systems,
it first {\it decreases} for increasing system sizes, before it increases
for larger systems and finally diverges in the macroscopic limit.
This behavious is in strong contrast to the treatment in the canonical ensemble
where a monotonic increase of the maximum of the specific heat with the size of the system
is always observed \cite{bind72,land76}. In \cite{behr05} 
a phenomenological theory has been proposed that acccounts
for the observed non-monotonic variation of the maximum of the microcanonical specific heat.

\subsection{Determination of critical exponents directly from the 
density of states}
At this stage one may wonder whether the typical features of spontaneous symmetry
breaking observed in finite microcanonical systems are described by the true critical
exponents of the infinite system. A rapid inspection of the numerical data reveals that
this is not the case. Indeed, the onset of the order parameter or the divergence
of the suceptibility in finite microcanonical systems are in all cases
governed by the classical mean-field exponents \cite{kast00,huel02,behr04}.
This is readily understood by noting that the entropy
surface can be expanded in a Taylor series in the vicinity of
$e_{c,N}$ (for a discrete spin model a suitable continuous function
that represents the data has to be chosen). As a 
consequence one ends up with a Landau-like theory \cite{behr04} which for the microcanonical
entropy of a finite system is in principle exact, thus yielding classical values
for the critical exponents. In recent years we have developed different approaches
that enable us nevertheless to obtain the true  critical exponents by exclusively
analyzing the density of states of finite systems \cite{huel02,behr05}.

Before discussing these approaches for the spontaneous 
magnetization in some detail we have first to pause briefly
in order to recall that in the entropy formalism considered here critical exponents
often differ from their thermal counterparts. This is the case when the specific
heat diverges algebraically at the critical point of the infinite system, implying that the critical exponent $\alpha$,
governing the power-law behaviour of the canonical specific heat in the vicinity of the
critical point, is positive. It is then easily shown that the critical exponents
$x_\varepsilon$ appearing in the microcanonical analysis are related to the
thermal critical exponents $x$ by $x_\varepsilon=x/(1-\alpha)$ \cite{kast00}. However, when the
specific heat does not diverge but presents a cusp-like singularity as $\alpha < 0$,
we have $x=x_\varepsilon$ \cite{behr04b,rich05}. In the case of a
logarithmically diverging specific heat the thermal and microcanonical
exponents are also identical.

Let us now come back to the determination of the critical exponents. As already
mentioned we obtain classical values for the critical exponents when expanding
the finite-size quantities around the energy $e_{c,N}$. The spontaneous
magnetization, for example, displays a root singularity for energies
$e < e_{c,N}$ close to $e_{c,N}$:
\begin{equation} \label{eq:mexp}
m_{sp,N}(e)= A_N \left( e_{c,N} - e \right)^{1/2}
\end{equation}
where the amplitude $A_N$ depends on the system size. In the infinite system
one expands around the true critical point $e_c$, leading to the asymptotic
behaviour
\begin{equation} \label{eq:mexpinf}
m_{sp}(\varepsilon) = A \, \varepsilon^{\beta_\varepsilon}
\end{equation}
for small $\varepsilon $
with $\varepsilon = \frac{e_c - e}{e_c - e_g}$, where $e_g$ denotes the ground state
energy per degree of freedom. The critical exponent $\beta_\varepsilon$ is given by
$\beta_\varepsilon = \beta/(1-\alpha)$ with $\beta$ being the usual thermal critical
exponent characterizing the singularity of the spontaneous magnetization. Here we suppose that $\alpha \geq 0$, as it is the case for the Ising
model. 

Away from the critical point the spontaneous magnetization of the infinite system
is not any more given by the simple expression (\ref{eq:mexpinf}), as correction
terms get important. One way to deal with these correction terms is to analyse 
the energy dependent effective exponents
\begin{equation} \label{eq:beff}
\beta_{\varepsilon,eff}(\varepsilon) = \frac{\D \ln m_{sp}(\varepsilon)}{\D \ln \varepsilon}
\end{equation}
that yields the true critical exponent $\beta_\varepsilon$ in the limit $\varepsilon 
\longrightarrow 0$ \cite{huel02}.

In order to use Equation (\ref{eq:beff})
we need data free of finite-size effects. On the other hand one observes in
Figure \ref{fig_3} that there exists a energy interval for which the spontaneous
magnetizations computed in systems of different sizes are identical.
This suggest the following procedure for computing $\beta_\varepsilon$.
Calculate the spontaneous magnetization for a series of increasing system sizes $\left\{ N_j , j=1, \cdots ,  n \right\}$.
In the energy range where $N_i$ and $N_{i+1}$ yield the same exponent 
$\beta_{\varepsilon,eff}$, select this exponent for the infinite system, until finite-size
effects show up. Then plot the common exponent for system sizes $N_{i+1}$ and
$N_{i+2}$ until these begin to disagree and so on. This procedure is rather cumbersome
but ensures that data essentially free of finite-size effects are obtained.

The result of this approach \cite{huel02} for the two- and the three-dimensional Ising models
is shown in Figure \ref{fig_4}. For the three-dimensional case we observe that 
$\beta_{\varepsilon,eff}$ rapidly approaches a constant value that perfectly agrees with the
expected asymptotic value $\beta_\varepsilon = 0.367$. In two dimensions the agreement is not 
as perfect, even so a reasonable estimate of $\beta_\varepsilon$ 
is obtained by linearly extrapolating the existing data.
The situation is indeed more complicated in the two-dimensional Ising model, as one
is dealing here with the rather pathological case of a logarithmically diverging specific
heat with $\alpha = 0$. In fact, we expect for any model with $\alpha \neq 0$
a similar good agreement between the expected value of
$\beta_\varepsilon$ and the value obtained from linearly extrapolating $\beta_{eff,\varepsilon}$ as for
the three-dimenisonal Ising model. This expectation
has recently been confirmed for the three-dimensional $XY$ model \cite{rich05}.

\begin{figure}
\centerline{\epsfxsize=4.0in\ \epsfbox{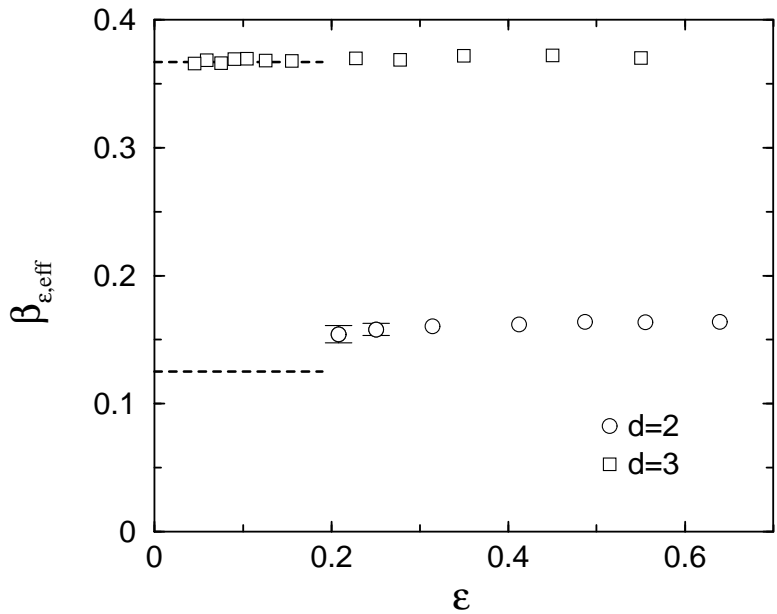}} 
\caption{Effective exponent $\beta_{\varepsilon,eff}$ versus reduced
energy $\varepsilon$ for the two- and the three-dimensional Ising models.
The dashed lines indicate the expected values of $\beta_\varepsilon$, see text.
Error bars are only shown when they are larger than the sizes of the symbols.} \label{fig_5}
\end{figure}

The determination of critical exponents via effective exponents has the major
drawback that data free of finite-size effects have to be obtained, which means
that systems of increasing sizes have to be simulated on approach to the critical 
point. For Figure \ref{fig_5} our largest systems contained $700 \times 700$ spins
in two and $80 \times 80 \times 80$ spins in three dimensions. Simulating these
systems is very time consuming even with the best algorithms available.
The second approach for determining the values of critical exponents discussed
in the following does not need large systems, but instead it explicitly takes
advantage of the finite-size effects.

In the canonical ensemble finite-size scaling theory is a valuable and often
used tool for extracting critical exponents from finite-size data. Its starting
point is the observation that in the asymptotic limit $L \longrightarrow \infty$ 
and $T \longrightarrow T_c$ ($T_c$ being the critical temperature)
the behaviour of finite-size quantities is governed
by scaling functions that are basically determined by the ratio
$L/\xi(T)$ with $\xi(T)$ being the correlation length and $L=N^{1/d}$
the linear extension of the $d$-dimensional system \cite{Bar83}.
In \cite{plei04} we have developed
a microcanonical finite-size scaling theory that takes advantage of
the existence of a well-defined transition point $e_{c,N}$ in finite
microcanonical systems.

\begin{figure}
\centerline{\epsfxsize=4.0in\ \epsfbox{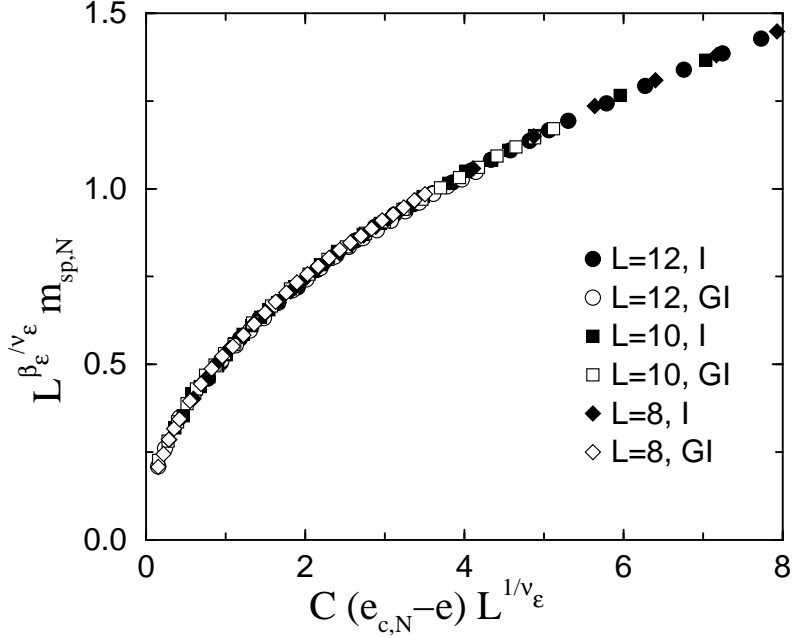}} 
\caption{Microcanonical finite-size scaling plot for the Ising model with only
nearest neighbour interactions (I) and for the Ising model with equivalent nearest and next-nearest
neighbour interactions (GI). Critical exponents are determined by the best data collapse.
By adjusting the nonuniversal constant $C$ the data of both models fall on a unique curve,
thus demonstrating the universality of the scaling function $W$ given in 
Eq.\ (\ref{eq:scal}). Error bars are smaller than the sizes of the symbols.} \label{fig_6}
\end{figure}

The starting point is the scaling behaviour of the entropy of finite systems
considered as a function of the energy, the magnetization and the inverse
system size. This then leads to the scaling form
\begin{equation}  \label{eq:scal}
L^{\beta_\varepsilon/\nu_\varepsilon} m_{sp,N}\left( e_{c,N} - e \right) 
\approx W\left( C ( e_{c,N} - e) L^{1/\nu_\varepsilon} \right)
\end{equation}
of the spontaneous magnetization (\ref{eq:msp}). Here $W$ is a scaling
function characterizing the given universality class and $C$ is a nonuniversal
constant which is different for the various model systems belonging to
the same universality class. The exponent $\nu_\varepsilon$ is given by $\nu_\varepsilon= \nu/(1-\alpha)$
where $\nu$ is the usual thermal critical exponent governing the divergence
of the correlation length on approach to the critical point.
As the spontaneous magnetization presents a square root singularity
in any finite microcanonical system, the scaling function $W$ varies for
small scaling variables $x=C ( e_{c,N} - e) L^{1/\nu_\varepsilon}$ as
$W(x) \sim \sqrt{x}$. In Figure \ref{fig_6} we test the scaling form (\ref{eq:scal})
for the usual Ising model (\ref{eq:ising}) with only nearest neighbour interactions
(denoted by I)
and for a generalized Ising model with equivalent nearest and next-nearest
neighbour interactions (denoted by GI). The Hamiltonian of this latter model is given by
\begin{equation} \label{eq:gi}
{\mathcal H}_{GI} = - \sum\limits_{\langle i,j \rangle} \, S_i S_j
- \sum\limits_{\left( i,k \right)} \, S_i S_k
\end{equation}
where the second sum is over bonds connecting next-nearest neighbours.
The model (\ref{eq:gi}) belongs to the same universality class as the standard
Ising model (\ref{eq:ising}) so that both models should yield the same values
for universal quantities.

Figure \ref{fig_6} illustrates two different aspects of the scaling behaviour of the
microcanonically defined spontaneous magnetization. On the one hand it shows that both
for the Ising model and for the GI model a data collapse can be achieved by plotting
$L^{\beta_\varepsilon/\nu_\varepsilon} m_{sp,N}$ as a function of 
$( e_{c,N} - e) L^{1/\nu_\varepsilon}$. The values of the involved critical exponents
are thereby obtained by the best data collapse. This yields the values
$\beta_\varepsilon/\nu_\varepsilon = 0.54 \pm 0.03$ ($0.51 \pm 0.03$) and
$1/\nu_\varepsilon = 1.43 \pm 0.01$ ($1.43 \pm 0.01$) for the Ising (GI) 
model, in excellent agreement
with the expected values $\beta_\varepsilon/\nu_\varepsilon = 0.52$ and
$1/\nu_\varepsilon = 1.43$. On the other hand Figure \ref{fig_6} also proves
that the scaling function $W$ appearing in Eq.\ (\ref{eq:scal}) is the same for both
models, in accordance with its expected universality. Indeed, the data of both models
fall on a common master curve when adjusting the nonuniversal constant $C$, so that
$C_{GI} = 0.219 C_I$. As shown in \cite{plei04} scaling functions obtained for models belonging
to different universality classes are indeed different.

The microcanonical finite-size scaling theory, which has also been
applied successfully to the
Potts \cite{plei04} and to the $XY$  \cite{rich05} models, is very promising. One of the remarkable
point is that no {\it a priori} knowledge of the infinite system is needed. Especially,
one does not need to know the exact location of the critical point of the infinite system.

\section{Conclusion}
In this paper we have discussed the signatures of phase transitions in finite
microcanonical systems. Discontinuous phase transitions are announced
in the microcanonical ensemble by a backbending of the caloric curve and the
appearance of negative specific heat. Continuous phase transitions also lead
to intriguing finite-size signatures. Indeed, typical features of spontaneous
symmetry breaking like the onset of the order parameter or the divergence of
the susceptibility are already encountered  in finite microcanonical systems.
These singularities are governed by the classical mean-field exponents. We have
discussed in this paper two different approaches that nevertheless enable us
to extract the true critical exponents from the density of states of finite systems.
One approach involves effective exponents, whereas the other is based on a
microcanonical finite-size scaling theory. The latter approach explicitly
takes advantage of the existence of a well-defined transition point in any
finite microcanonical system.\\~\\

{\bf Acknowledgements:}\\~\\
It is a pleasure to thank Alfred H\"{u}ller for many years of fruitful
and enjoyable collaboration. 

\end{document}